\documentclass[preprint,showpacs,pra,aps]{revtex4}

\usepackage{graphicx}

\begin{document}

\title{Enrichment of CH$_3$F nuclear spin isomers 
       by\\ resonant microwave radiation}
       
\author{O.~I.~Permyakova}

\affiliation{Institute of Semiconductor Physics, 
Russian Academy of Sciences, 630090 Novosibirsk, Russia}
          
\author{E.~Ilisca}

\affiliation{Laboratoire de Physique Th\'{e}orique de la Mati\`{e}re 
Condens\'{e}e, Universit\'{e} Paris 7--Denis Diderot,\\ 
2, Place Jussieu, 75251 Paris Cedex 05, FRANCE}
         
\author{P.~L.~Chapovsky\thanks{E-mail: chapovsky@iae.nsk.su}}
\email[E-mail: ]{chapovsky@iae.nsk.su}

\affiliation{  Institute of Automation and Electrometry,
          Russian Academy of Sciences, 630090 Novosibirsk, Russia}
         

\begin{abstract}
Theoretical model of the coherent control of nuclear spin isomers 
by microwave radiation has been developed. Model accounts the 
$M$-degeneracy of molecular states and molecular center-of-mass motion. 
The model has been applied to the $^{13}$CH$_3$F molecules. 
Microwave radiation excites the para state ($J$=11,$K$=1) which is mixed by
the nuclear spin-spin interaction with the ortho state (9,3).
Dependencies of the isomer enrichment and conversion rates on the radiation
frequency have been calculated. Both spectra consist of two resonances 
situated at the centers of allowed and forbidden (by nuclear spin)
transitions in the molecule. Larger enrichment, up to 7\%, can be 
produced by strong radiation resonant to the forbidden
transition. The spin conversion rate can be increased by 2 orders of 
magnitude at this resonance. 
\end{abstract}


\maketitle

\section{Introduction}

Nuclear spin isomers of symmetrical molecules
are fascinating objects \cite{Landau81}. Their properties are determined 
by nuclei' spins deeply hidden in the molecule. Most known 
are the hydrogen isomers that demonstrate anomalous 
stability, 1 year at ambient temperature and pressure \cite{Farkas35}. Latest
results on hydrogen isomers can be found in \cite{Ilisca92,Ilisca99PRL} 
and references therein. Many other molecules have spin isomers too. But
so far their physical properties remain almost unknown. This makes investigations 
of spin isomers fundamentally important. 
Spin isomers have also practical applications, e.g., as spin labels, 
in isomer selective chemical reactions \cite{Quack77MP,Uy97PRL}, or in isomer
enhanced NMR technique \cite{Bowers86PRL,Natterer97PNMRS}.
These applications are developed solely with hydrogen isomers.
Extension to other molecules needs efficient methods of isomer enrichment.
For a long time enrichment of only hydrogen isomers was possible.  
Recently a few separation methods for polyatomic molecules 
have been developed (see the review \cite{Chap99ARPC}) that has advanced the 
field significantly. Further progress needs new enrichment methods. 

New approach to the problem of isomer enrichment is 
based on the use of strong electromagnetic radiation.
There are two modifications of the method. In the first one
\cite{Ilichov98CPL,Shalagin99JETPL} (earlier discussion of the radiation
effects see in \cite{Ilisca88CPL}),
radiation populates specific states of one spin isomer situated 
in the vicinity of states of other isomer that 
consequently results in the enrichment. In the second
method \cite{Chap01PRA2}, enrichment appears due to combined action of 
population transfer, dynamical shift of molecular levels
and light induced coherence. The latter method 
(coherent control of spin isomers) promises to be more efficient.
To avoid any confusion, we note that the light-induced enrichment
resulting from stimulating conversion of spin species differs radically
from the previously known separation methods, e.g., light-induced
drift method which separates physically the isomers \cite{Chap99ARPC}.

There are no proofs yet that light-induced enrichment of spin isomers 
is feasible. We are aware of only one attempt to verify the proposals.
It was done by microwave excitation of rotational transition in CH$_3$F. 
The experiment did not give a positive result.
At the time when this experiment was performed, only a qualitative 
theoretical model of the light-induced enrichment was available
\cite{Ilichov98CPL}. Presently, the understanding of the underlying physics 
has been improved substantially. In view of further experiments in this area
it is desirable to examine the microwave induced
enrichment of CH$_3$F spin isomers in more detail. This is the goal of the present paper. 
Existing theoretical models of coherent control cannot be used for quantitative analysis
directly because of their  lacking to account the $M$-degeneracy of molecular states. 
Account of such degeneracy is another goal of this paper.

\section{Qualitative picture and kinetic equation}

Previous analysis has shown that significant enrichment can be obtained
if radiation interacts with the states having large difference
in populations. In this context, microwave excitation is not the best 
option but it has some advantages also. Theoretical description
is simpler for pure rotational excitation. The levels positions are 
better known for ground vibrational states.  From the experimental
side, it is easier to find a radiation having proper frequency
because of better frequency tunability of microwave sources. 
 
We start with brief qualitative description of the phenomenon. CH$_3$F has  
two types of states, ortho and para, shown in Fig.~\ref{fig1}.
The data in this figure correspond to the 
$^{13}$CH$_3$F molecule and have been calculated using the molecular 
parameters from \cite{Papousek94JMS}. Spin isomers of CH$_3$F are
distinguished by the total spin of the three hydrogen nuclei,
$I=3/2$ for ortho and $I=1/2$ for para isomers. For ortho isomers
only rotational quantum numbers $K=0, 3, 6..$ are allowed ($K$ is
the projection of molecular angular momentum, {\bf J}, on the molecular symmetry
axis.) For para molecules only $K=1, 2, 4, 5...$ are allowed
\cite{Landau81}.

There are two close pairs of ortho and para states in the ground 
vibrational state of $^{13}$CH$_3$F that are significantly mixed by 
the intramolecular perturbation, $\hat V$, and that are 
important for the ortho-para conversion
in the molecule. For a qualitative description, let us take into 
account only one of these pairs, $m-n$, and assume that there
is no external radiation yet.
Suppose that the test molecule is placed into the ortho subspace.
Due to the rotational relaxation caused by collisions, the 
molecule starts to shuttle up and down along the ladder
of rotational states. 
Nonmagnetic collisions do not change the
nuclear spin state directly, i.e., the relevant cross-section is zero,
$\sigma(ortho|para)=0$. This shuttling along the rotational states inside the ortho
subspace continues until the molecule jumps to
the state $m$. During the free flight after that collision 
the intramolecular perturbation, $\hat V$,   
admixes the para states $n$ to the ortho state $m$.  Consequently,
the next collision has a probability (usually very small) to 
transfer the molecule to other 
para states. This localizes the molecule inside the para subspace and 
the spin conversion occurs. This is the mechanism of radiation 
free nuclear spin conversion induced by the intramolecular state
mixing \cite{Curl67JCP} (see also \cite{Chap91PRA}).

In case of a strong microwave radiation applied to the molecular
transition $q-n$ in the para subspace, mixing of the states
is affected by the radiation which allows to 
control the ortho-para conversion. Influence of a radiation
comes through the three major effects, level shift (dynamical Stark 
effect, well-known in nonlinear laser spectroscopy 
\cite{Rautian79,Cohen-Tannoudji92}), level population change, 
and light-induced coherence. In general, these three components cannot 
be separated and work together. 

In order to consider a real molecule, the above simplified picture 
has to be developed further. One has to account the molecular center-of-mass motion.
Although the intramolecular mixing does not depend on molecular velocity,
the radiation-molecular interaction does. Consequently, the ortho-para
state mixing in coherent control depends on molecular velocity too.

Another complication comes from the degeneracy 
of molecular states. Even for the simplest case of pure radiation
polarization (linear or circular) there are many excitation channels 
each having its own absorption coefficient
and saturation parameter. These channels differ by $M$-quantum
number, projection of {\bf J} on the laboratory axis of quantization.
It is important also to keep in mind that there are other degeneracies
of states. Each state of $^{13}$CH$_3$F in Fig.~\ref{fig1} 
is determined by the set of rotational quantum numbers ($J,K,M$), 
total spin of three hydrogen, $I$, its projections on the laboratory 
$z$-axis,  $\sigma$, and $z$-projects of spins of carbon and fluorine nuclei,
both having spin 1/2. The energy of rotational states of CH$_3$F depends
only on $J$ and $K$ quantum numbers, if tiny hyperfine contribution
to the level energy is neglected.     
We end the qualitative picture by summarizing important parameters of 
$^{13}$CH$_3$F in Table~1.

Quantitative analysis of the isomer coherent control will be performed 
using kinetic equation for the density matrix, $\hat\rho$.
The molecular Hamiltonian reads,
\begin{equation}
     \hat H = -(\hbar^2/2m_0)\nabla^2_{\bf r} + \hat H_0 + 
              \hbar\hat G + \hbar\hat V.
\label{H}
\end{equation} 
Here the first term is the Hamiltonian of the molecular center-of-mass motion
with $m_0$ being the molecular mass.
The main part of the molecular internal Hamiltonian, $\hat H_0$, has 
the eigen ortho and para states 
shown in Fig.~\ref{fig1}. $\hbar\hat G$ describes 
the molecular interactions with the external radiation that will be
taken in the form of monochromatic travelling wave,
\begin{equation}
\hat G = - ({\bf E}_0\hat{\bf d}/\hbar)\cos(\omega_Lt-{\bf kr}),   
\label{G}
\end{equation}
where ${\bf E}_0$, $\omega_L$ and $\bf k$ are the amplitude, frequency 
and wave vector of the electromagnetic radiation, respectively; $\hat {\bf d}$ 
is the operator of the molecular electric dipole moment. 
$\hat V$ is the intramolecular perturbation
that mixes the ortho and para states in $^{13}$CH$_3$F. The mixing 
of $m-n$ pair ($J$=9,$K$=3--11,1) is performed by the
spin-spin interaction between the molecular nuclei \cite{Chap91PRA,Chap00EPJD}, 
The pair $m'-n'$ (20,3--21,1) is mixed by the 
spin-spin and spin-rotation interactions 
\cite{Chap91PRA,Guskov99JPB,Ilisca98PRA,Cacciani02}.
Account of the level degeneracy for the light-molecular interaction is
a difficult problem, in general. It is most simple 
for the case of pure polarization, linear or circular.
We will consider the electromagnetic radiation having linear polarization.

In the representation of the eigen states of $\hat H_0$ 
($\alpha$-states) and classical description
of the molecular center-of-mass motion, kinetic equation reads
\cite{Rautian79},
\begin{equation}
    \partial\text{\boldmath$\rho$}/\partial t +{\bf v}\cdot\nabla\text{\boldmath$\rho$} = 
    {\bf S} - i [{\bf G}+{\bf V},\text{\boldmath$\rho$}]. 
\label{r1}
\end{equation}
Here ${\bf S}$ is the collision integral; ${\bf v}$ is
the molecular center-of-mass velocity. Spontaneous decay is not included 
in this equation because it is negligible for rotational transitions 
in comparison with the collisional relaxation. 

Kinetic equation for the total concentration of para molecules 
can be obtained directly from Eq.~(\ref{r1}), 
\begin{equation}
     \partial\rho_p/\partial t = 
     -2Re \int i(\sum\rho_{m'n'}V_{n'm'}+\sum\rho_{mn}V_{nm})d{\bf v}.
\label{ro}
\end{equation} 
Here the total concentration of para molecules, 
$\rho_p = \sum_\alpha\int\rho_p(\alpha,{\bf v})d{\bf v},\ \ \alpha\in$~para.
Summation is made over all degenerate sublevels of the states 
$m',n'$ and $m,n$. In Eq.~(\ref{ro})
a uniform spatial distribution of molecular density was assumed. 
Collision integral did not enter into Eq.~(\ref{ro}) because by assumption 
collisions do not change the molecular spin state, i.e.,
$\sum_\alpha\int S_{\alpha\alpha}d{\bf v}=0$, if $\alpha\in$~ortho, 
or $\alpha\in$~para. {\bf G} did
not enter into Eq.~(\ref{ro}) either because the matrix elements of {\bf G}   
off-diagonal in nuclear spin states vanish.    

The off-diagonal matrix elements $\rho_{mn}$ and $\rho_{m'n'}$ will be found
in perturbation theory. Further we assume the perturbations $\hat V$
being small and consider zero- and first-order terms of the 
density matrix,
\begin{equation}
     \text{\boldmath$\rho$} = \text{\boldmath$\rho$}' + 
                              \text{\boldmath$\rho$}''.
\label{split}
\end{equation}

Collisions in our system will be described by the model standard
in the theory of light-molecule interaction. The off-diagonal elements 
of ${\bf S}$  have only decay terms,
\begin{equation}
     S_{\alpha\alpha'} = -\Gamma\rho_{\alpha\alpha'}; \ \ \  \alpha\neq \alpha'.
\label{S_off}
\end{equation}
The decoherence rates, $\Gamma$, are taken equal for all 
off-diagonal elements of collision integral. This assumption simplifies
the theoretical model. Note, that the dependence
of the relaxation rates on rotational quantum numbers is known
\cite{Trappeniers79JCP}.

The diagonal elements of ${\bf S}$ are expressed 
through the kernel of collision integral, $A$, in a usual way,
\begin{equation}
   S(\alpha,{\bf v}) = \sum_{\alpha_1}\int A(\alpha,{\bf v}|\alpha_1,{\bf v}_1)
   \rho(\alpha_1,{\bf v}_1)d{\bf v}_1 -
   \rho(\alpha,{\bf v})\sum_{\alpha_1}\int A(\alpha_1,{\bf v}_1|\alpha,{\bf v})d{\bf v}_1.
\label{S}
\end{equation}
We consider the model of strong collisions with the following collision kernel
for para molecules
\begin{equation}
     A(\alpha,{\bf v}|\alpha_1,{\bf v_1}) = \nu_rw_p(\alpha)\delta({\bf v}-{\bf v_1}) +
     \nu_t\delta_{\alpha\alpha_1}f({\bf v});\ \ \ \alpha,\alpha_1\in {\text para},
\label{A}
\end{equation}
and similar equation for ortho molecules. $w_p(\alpha)$ in Eq.~(\ref{A}) is the Boltzmann 
distribution of rotational state populations of para molecules,
\begin{equation}
     w_p(\alpha)=Z^{-1}_p\exp(-E_\alpha/k_BT),
\label{B}
\end{equation}
with $Z_p$ being the rotational partition function;
$E_\alpha$ the rotational energy of $\alpha$-state; $T$ the
gas temperature; $k_B$ the Boltzmann constant.  
The symmetry of CH$_3$F is such that the partition functions for ortho 
and para molecules are practically equal at room temperature.
Partition functions account all degeneracies of states 
(see Ref.~\cite{Chap91PRA} for more details); $f({\bf v})$ in Eq.~(\ref{A}) 
is the Maxwell distribution,
\begin{equation}
     f({\bf v}) =\pi^{-3/2}v_0^{-3}\exp(-{\bf v}^2/v^2_0);\ \ \ 
     v_0=\sqrt{2k_BT/m_0}. 
\label{M}
\end{equation}

In Eq.~(\ref{A}), two relaxation rates were introduced, rotational 
relaxation ($\nu_r$) that does not affect molecular velocity and translational relaxation 
($\nu_t$) that equilibrates velocity but do not change rotational state.
Note that the rotational relaxation
is accompanied in our model by the relaxation in $M$ quantum numbers. 
Note also, that the collisions in the model do not change the molecular spin
state. The introduction of different relaxation 
rates for different degrees of freedom
makes the model of strong collisions more accurate and flexible. It allows
to adjust the model to the particular experimental conditions. 
Because of its simplicity model of strong collisions is often used in 
laser physics and nonlinear spectroscopy, see, e.g., 
\cite{Rautian79,Dykhne80JETP,Mironenko81IANS}. Numerical values for the
collisional parameters $\Gamma, \nu_r$, and $\nu_t$ will be determined later.

\section{Microwave absorption}

For the zero-order term of the density matrix one has the following kinetic 
equation,
\begin{equation}
      \partial\text{\boldmath$\rho$}'/\partial t +
      {\bf v}\cdot\nabla\text{\boldmath$\rho$}' = 
      {\bf S}'  - i [{\bf G},\text{\boldmath$\rho$}'].
\label{ro0}
\end{equation}
Electromagnetic field interacts with para molecules only. Consequently, ortho
molecules remain at equilibrium in the zero-order perturbation theory. 
For the level populations of ortho molecules one has,
\begin{equation}
     \rho'_o(\alpha,{\bf v}) = 
     (N-\rho'_p) w_o(\alpha)f({\bf v}), 
\label{0rp}
\end{equation}
where $N$ is the total concentration of molecules.

Eqs.~(\ref{S_off}),(\ref{S}),(\ref{A}), and (\ref{ro0}) allow to 
deduce an equation for the stationary populations of para molecules,
\begin{equation}
(\nu_r+\nu_t)\rho'_p(\alpha,{\bf v})  =  \nu_r w_p(\alpha)\rho'_p({\bf v})
         + \nu_t f({\bf v})\rho'_p(\alpha)    
         + \rho'_p p [\delta_{\alpha q}-\delta_{\alpha n}],
\label{rhop}
\end{equation}
where the excitation probability, $p$, is defined as,
\begin{equation}
 \rho'_p p = \frac{2\Gamma |G_{qn}|^2}{\Gamma^2 + (\Omega-{\bf kv})^2}
                \left[\rho'_p(n,{\bf v})-\rho'_p(q,{\bf v})\right].      
\label{p}
\end{equation}
In Eq.~(\ref{rhop}) the notations were introduced,
\begin{equation}
     \rho'_p({\bf v}) = \sum_{\alpha\in para} \rho'_p(\alpha,{\bf v});\ \ \  
     \rho'_p(\alpha) = \int \rho'_p(\alpha,{\bf v}) d{\bf v}.
\label{rr}
\end{equation} 
Eq.~(\ref{p}) is written in the rotating wave approximation.
Nonzero matrix elements of ${\bf G}$ (electric field has linear polarization along
$z$-axis) are given by,
\begin{equation}
     G_{qn} = G(M)e^{i({\bf kr}-\Omega t)};\ \ 
     G(M)\equiv E_{10} (\overline{d_{10}})_{qn}/2\hbar, 
\label{rwa}
\end{equation}
where $\Omega=\omega_L - \omega_{qn}$ is the radiation frequency
detuning from the absorption line center, $\omega_{qn}$; 
the bar over a symbol indicates a time-independent factor; 
$E_{10}$ and $d_{10}$ are spherical components of the electric field and
electric dipole moment vectors, respectively \cite{Landau81}.  
The matrix elements of $\hat d_{10}$ reads \cite{Landau81},
\begin{equation}
     |(\overline{d_{10}})_{qn}|^2 \equiv |d(M)|^2 = (2J_q+1)(2J_n+1)
     \left(\begin{array}{ccr}
                    J_q & 1 & J_n \\ 
                     -K & 0 & K
                \end{array}\right)^2
                \left(\begin{array}{ccr}
                    J_q & 1 & J_n \\ 
                     -M & 0 & M
                \end{array}\right)^2d^2,
\label{d}
\end{equation}
where (:::) stands for the 3j-symbol and $d$ is the permanent electric 
dipole moment of CH$_3$F, $d=1.86$~D \cite{Freund74JMS}.

Solution of Eq.~(\ref{rhop}) has no difficulty and can be presented in the form,
\begin{equation}
     \rho'_p(\alpha,{\bf v}) = \rho'_p w_p(\alpha) f({\bf v}) +
                     \rho'_p[(\tau_2-\tau_1)p_1f({\bf v}) + 
                     \tau_1p][\delta_{\alpha q}-\delta_{\alpha n}],
\label{rho1}
\end{equation}
where the relaxation times are $\tau_1=(\nu_r+\nu_t)^{-1}$, $\tau_2=\nu_r^{-1}$, 
and $p_1=\int pd{\bf v}$. We have separated here the field-induced contributions
nonequilibrium in $\alpha$ and {\bf v} and nonequilibrium only in $\alpha$. 
Solution (\ref{rho1}) shows that radiation affects the population of 
only two states, $q$ and $n$. This is the consequence
of the accepted simple model of collisions.

The excitation probability can be found from Eqs.~(\ref{p}) and (\ref{rho1}),
\begin{eqnarray}
p_1 & = & \frac{0.5\Delta w}{\tau_1(\kappa R)^{-1}+
                   \tau_2-\tau_1}, \nonumber \\
  p & = & \frac{\Gamma^2f({\bf v})}{\Gamma^2_B+(\Omega-{\bf kv})^2}
                    \frac{p_1}{R},       
\label{pp}
\end{eqnarray}
where the difference of the Boltzmann factors is $\Delta w = w_p(n)-w_p(q)$;
the saturation parameter, $\kappa$, and saturation intensity, $S_{sat}$,
are
\begin{equation}
      \kappa = \frac{S}{S_{sat}}, \ \ \ 
     S_{sat} = \frac{c \Gamma\hbar^2}{8\pi\tau_1|d(M)|^2};
\label{ki}
\end{equation}
the homogeneous line width is $\Gamma_B=\Gamma\sqrt{1+\kappa}$, and
\begin{equation}
     R = \int \frac{\Gamma^2f({\bf v})d{\bf v}}{\Gamma^2_B+(\Omega-{\bf kv})^2}.
\label{R}
\end{equation}
This integral can be expressed through the probability integral, but for
numerical calculations performed in this paper it is easier to calculate
it straightforward.

In a similar way one can obtain from the 
kinetic equation (\ref{ro0}) the off-diagonal density matrix element,
\begin{equation}
     \overline\rho'_{qn} = -i\frac{\rho'_p}{\overline{G}_{nq}}
                 \frac{p}{2\Gamma}[\Gamma + i(\Omega-{\bf{kv}})].
\label{ronk}
\end{equation}

We can adjust now parameters of the collision kernel (\ref{A}). 
Kinetic equation (\ref{ro0}) describes a diffusion process with the 
diffusion coefficient, $D=v^2_0/2\nu_t$. Diffusion coefficient for CH$_3$F 
is equal to $D\simeq10^2$~cm$^2$/s at the pressure 1~Torr. This determines 
the velocity equilibration rate, $\nu_t=4.4\cdot10^7$~s$^{-1}$/Torr.

Attenuation of the radiation is given by $\hbar\omega_L\rho_p\sum_Mp_1$.
Consequently the absorption coefficient, $\chi(\Omega)$, is determined by 
the expression,
\begin{equation}
     \chi(\Omega) = \hbar\omega_L\rho_pS^{-1}\sum_M p_1.
\label{chi}
\end{equation}
In the low field limit ($\kappa\rightarrow 0$) it is reduced to,
\begin{equation}
     \chi_{low}(\Omega) = \frac{\hbar\omega_L}{2\tau_1}\rho_p\Delta w
     R_{low}\sum_M S_{sat}^{-1},
\label{chilow}
\end{equation}
where $R_{low}=\lim_{S\rightarrow0} R$ is the Foigt profile of the
absorption line. If $\Gamma\gg kv_0$ the absorption line is Lorentzian
having the width equal $\Gamma$. Experimental data on $\chi(\Omega)$
for the rotational transition $11,1\rightarrow12,1$ can be used to determine 
the value of $\Gamma$. Equally, broadening of any other rotational transition 
can be used because in our collision model (\ref{S_off}) one has the same
$\Gamma$ for all off-diagonal density matrix elements. There are
experimental results on the broadening of the ortho-para transition 
(9,3)--(11,1) obtained from
the level-crossing resonances in $^{13}$CH$_3$F nuclear spin conversion.
This experiment gave the value $\Gamma/P=1.9\cdot10^8$~s$^{-1}$/Torr 
\cite{Nagels96PRL,Chap00AMR} that will be used in the present calculations.
The last unknown parameter, rotational relaxation, $\nu_r$, can be 
determined, e.g., from the power saturation of the absorption coefficient.
This information is not available and we assume $\nu_r=\Gamma$. This is
reasonable, because the pressure broadening in molecules is determined 
mainly by the level population quenching, although this estimation
for $\nu_r$ is probably too high. 

We can now demonstrate the model at work by considering the microwave 
absorption by $^{13}$CH$_3$F. Absorption spectrum is determined by the
selection rules $J\rightarrow J+1$, $K\rightarrow K$. The spectrum consists
of groups of lines nearly equally separated by 50~GHz. Inside each 
group the lines, different in $K$, are rather dense. Spectrum near the line
$11,1\rightarrow12,1$ is shown in Fig.~\ref{fig2}. The two spectra correspond
to low radiation intensity and to $S=100$~mW/cm$^2$ and the gas pressure 
equal 30~mTorr in both cases.

Saturation intensity for the line $11,1\rightarrow12,1$ is equal to 
43~W/cm$^2$ ($M=11$) and  6.8~W/cm$^2$ 
($M=0$) at the gas pressure 1~Torr. $S_{sat}$ is
proportional to the pressure squared, thus at 30~mTorr, $S_{sat}\simeq$6~mW/cm$^2$.
An example of the absorption coefficient saturation is given in Fig.~\ref{fig3}.
Because the Doppler width of the transition is small, $kv_0=0.74$~MHz,
low field absorption in the line center  
depends weakly on CH$_3$F pressure if $P\gtrsim100$~mTorr.
Another example of the saturation effect is shown in Fig.~\ref{fig4}. Here the 
relative level population difference $[\rho_p(n)-\rho_p(q)]/\rho_p\Delta w$ is given
as a function of radiation intensity. One can see that radiation
having $S=100$~mW/cm$^2$ 
decreases the level population difference significantly.

\section{First order theory}

The kinetic equation for the first order term of the density matrix
$\text{\boldmath$\rho$}''$ is obtained from Eq.~(\ref{r1}), 
\begin{equation}
     \partial\text{\boldmath$\rho$}''/\partial t + 
     {\bf v}\cdot\nabla\text{\boldmath$\rho$}'' = {\bf S}'' 
        - i [{\bf G},\text{\boldmath$\rho$}''] 
        - i [{\bf V},\text{\boldmath$\rho$}'].
\label{ro1}
\end{equation}
Ortho-para conversion is determined by the terms $\rho''_{m'n'}$ and 
$\rho''_{mn}$ because zero order matrix elements off-diagonal in
nuclear spins vanish (see Eq.~(\ref{ro})). Radiation does not 
affect the levels $m'$ and $n'$.
Consequently, the matrix element $\rho''_{m'n'}$ is not different
from the case of the field free conversion \cite{Chap91PRA},
\begin{equation}
     \rho''_{m'n'} = \frac{-iV_{m'n'}}{\Gamma + i\omega'}
     [\rho'_p(n',{\bf v})-\rho'_o(m',{\bf v})],
\label{rm'n'}
\end{equation}
where $\omega'\equiv\omega_{m'n'}$. The density matrix element
$\rho''_{mn}$ can be obtained from the equations
which are deduced from Eq.~(\ref{ro1}),
\begin{eqnarray}
(\partial/\partial t + {\bf v}\cdot\nabla\ +  \Gamma)\rho''_{mn} - i\rho''_{mq}G_{qn}
          & = & -iV_{mn}[\rho'_p(n,{\bf v})-\rho'_o(m,{\bf v})]; \nonumber \\
(\partial/\partial t  + {\bf v}\cdot\nabla + \Gamma)\rho''_{mq} - i\rho''_{mn}G_{nq} 
          & = & -iV_{mn} \rho'_{nq}.    
\label{sys1}
\end{eqnarray}
Substitutions, 
     $V_{mn}=\overline{V}_{mn}e^{i\omega t},\ \ (\omega\equiv\omega_{mn});\ \ 
     \rho''_{mn} = \overline{\rho}''_{mn}e^{i\omega t};\ \ 
     \rho''_{mq} = \overline{\rho}''_{mq}
     e^{i[(\Omega+\omega)t-{\bf kr}]}$,
transform Eqs.~(\ref{sys1}) to algebraic equations which can be easily solved.
The density matrix element that one needs for the kinetic equation 
(\ref{ro}) reads,
\begin{equation}
\overline\rho''_{mn}  =  -i\overline V_{mn}
   \frac{[\Gamma+i(\Omega+\omega-{\bf kv})][\rho'_p(n,{\bf v})-
           \rho'_o(m,{\bf v})]+i\overline G_{qn}\overline\rho'_{nq}}
           {(\Gamma+i\omega)[\Gamma+i(\Omega+\omega-{\bf kv})]+|G(M)|^2}.       
\label{rmn}
\end{equation}
Note that indices $m, n, q$ in Eqs.~(\ref{rmn}) represent the set
of quantum numbers that comprise all degenerate sublevels. Consequently, one has 
nonzero terms $\overline\rho''_{mn}$ for the combination of quantum numbers which 
are allowed by the selection rules for $\hat V$.

\section{Conversion rates}

Solutions for $\overline\rho''_{m'n'}$ and $\overline\rho''_{mn}$ together
with the level populations from Eq.~(\ref{rho1}) and the off-diagonal matrix
element from Eq.~(\ref{ronk}) allow to present Eq.~(\ref{ro}) as,
\begin{equation}
     \partial\rho_p/\partial t = N(\gamma_{po}'+\gamma_{po})
        -\rho_p\gamma;\ \ \ 
        \gamma\equiv\gamma_{po}'+\gamma_{op}'+\gamma_{po}+\gamma_{op}-\gamma_n-\gamma_c.
\label{drp1}
\end{equation}
In this equation we have neglected the small difference between the total concentration
of para molecules, $\rho_p$, and its zero order approximation, $\rho'_p$.
The partial conversion rates in Eq.~(\ref{drp1}) have the following definition. 
The field free conversion rate through the upper level pair, $m'-n'$,
\begin{equation}
     \gamma_{po}'=\sum \frac{2\Gamma|V_{m'n'}|^2}{\Gamma^2+\omega'^2}w_o(m').
\label{g'}
\end{equation}
Equation for $\gamma_{op}'$ is obtained from $\gamma_{po}'$ by
substitution the Boltzmann factor $w_p(n')$ instead of $w_o(m')$.
Summation is made here over all degenerate substates of $m'$ and $n'$ states.
The rate $\gamma_{po}$ is given by,
\begin{equation}
    \gamma_{po} = \sum |V_{mn}|^2 
      \left[\frac{2\Gamma}{\Gamma^2+\omega^2}+Re\int F_1f({\bf v})d{\bf v}\right]w_o(m)
\label{g}
\end{equation}
Equation for $\gamma_{op}$ is obtained from $\gamma_{po}$ by 
substitution the Boltzmann factor $w_p(n)$ instead of $w_o(m)$.
The rates $\gamma_{po}$ and $\gamma_{op}$ are field dependent.
Their zero field limits coincide with the field free conversion rates 
through the pair of states $m-n$.
The ``non-coherent'' contribution to the conversion, the rate $\gamma_n$, 
originated from the radiation-induced level population change 
in Eq.~(\ref{rmn}), is given by,
\begin{equation}
     \gamma_n = \sum |V_{mn}|^2\left[\frac{2\Gamma\tau_2p_1}{\Gamma^2+\omega^2}+
                2Re\int[(\tau_2-\tau_1)p_1f({\bf v}) + \tau_1 p]F_1d{\bf v}\right].
\label{gn}
\end{equation}
And finally, the ``coherent'' contribution to the conversion rate,
$\gamma_c$, originated from the $\overline\rho'_{nq}$
in Eq.~(\ref{rmn}) is,
\begin{equation}
     \gamma_c = \sum |V_{mn}|^2\left[-\frac{p_1}{\Gamma^2+\omega^2}
                 +\frac{1}{\Gamma}Re\int pF_2d{\bf v}\right]
\label{gc}
\end{equation}
In Eqs.~(\ref{g}),(\ref{gn}), and (\ref{gc}) the following functions were introduced,
\begin{eqnarray}
F_1 & = & \left(1-\frac{\Gamma_1+i\omega_1}{\Gamma+i\omega}\right)
          \frac{1}{\Gamma_1+i(\Omega+\omega_1-{\bf kv})}; \ \ \ 
          \Gamma_1=\Gamma\left(1+\frac{|G(M)|^2}{\Gamma^2+\omega^2}\right); \nonumber \\
F_2 & = & \frac{\Gamma+\Gamma_1+i\omega_1}{\Gamma+i\omega}
          \frac{1}{\Gamma_1+i(\Omega+\omega_1-{\bf kv})};\ \ \ 
          \omega_1=\omega\left(1-\frac{|G(M)|^2}{\Gamma^2+\omega^2}\right).      
\label{f1}
\end{eqnarray}

Interpretation of the introduced conversion rates can be considered as
follows. Strong resonant radiation splits the molecular states, change 
the levels populations and introduce coherences in the molecule.  
The field dependent part of $\gamma_{po}$ can be considered as
due to the radiation induced level crossing. This term has
resonance at $\Omega=-\omega_1$. The first term of $\gamma_n$ is due to the
population effect. It has the resonance at $\Omega=0$ where the
excitation probability has maximum. The second part
of $\gamma_n$ is due to the population change and level-crossing. It has
two resonances, at $\Omega=0$  and at $\Omega=-\omega_1$. The coherent
contribution, $\gamma_c$, has also resonances at these two frequencies.

The two peaks in the conversion rate spectra have rather distinctive features.
The first one, at $\Omega=0$,  is quite similar to the radiation 
free conversion rate, e.g., $\gamma'_{po}$ (Eq.~(\ref{g'})). 
In our case $\omega\gg\Gamma$ and the amplitudes of the peaks at 
$\Omega=0$ are proportional to $\Gamma$, thus to the gas
pressure. In the limit $\omega\gg\Gamma$, contributions to the 
conversion rate provided by these
terms are similar to the ordinary gas kinetic processes which are
proportional to the gas pressure also.

The resonances at $\Omega=\omega_1$ have completely different signature 
that would result from a field free conversion pattern of a degenerate
ortho-para level pair. In this case the conversion rate has 
$1/\Gamma$ dependence, see, e.g., 
Eq.~(\ref{g'}). It allows us to refer to the resonances at 
$\Omega=\omega_1$ as produced by the crossing the ortho and para levels
and resulting from the applied electromagnetic field. To reveal 
this property of the resonances at $\Omega=-\omega_1$, let us estimate
the integral,
\begin{equation}
     I = \int \frac{f({\bf v})d{\bf v}}{\Gamma_1+i(\Omega+\omega_1-{\bf kv})},
\label{if1}
\end{equation}
at the radiation frequency $\Omega=\omega_1$. If the limit
of large Doppler broadening is valid, $\Gamma_1\ll kv_0$, the integrand in 
Eq.~(\ref{if1}) has sharp resonance at ${\bf v}=0$. 
One can substitute $f({\bf v})$ by $f(0)$ in Eq.~(\ref{if1}) and obtain, 
$I\propto1/kv_0$. Thus, the conversion is produced through
the degenerate ortho-para level pair (level crossing) having the width
equal the Doppler width, $kv_0$.
In the opposite limit of large homogeneous broadening, $\Gamma_1\gg kv_0$, one can neglect 
${\bf kv}$ in the denominator of the integrand (\ref{if1})
and obtain $I\propto1/\Gamma_1$. Again, this is the conversion through 
the crossed ortho and para states but having the width
$\Gamma_1$ now.

It would be useful to compare the results of the present model
with the qualitative model \cite{Ilichov98CPL}. This comparison cannot be
made directly because in \cite{Ilichov98CPL} rovibrational excitation of
CH$_3$F was considered.
But one can apply the idea of light-induced enrichment solely through 
the level population change and make the comparison.
In the present notations, the field effect from the level population change
is given by the first term in Eq.~(\ref{gn}),
\begin{equation}
     \gamma_{n1} = \sum \frac{2\Gamma|V_{mn}|^2}{\Gamma^2+\omega^2}\tau_2p_1.
\label{gn1}
\end{equation}
It gives larger amplitude for the peak at $\Omega=0$ than the present
model. In the present model one has partial cancellation of peaks 
at $\Omega=0$, see the first terms in the expressions for $\gamma_n$
and $\gamma_c$. Because of the assumption, $\Gamma=\nu_r$,
made in the collision model we have two times smaller peak at 
$\Omega=0$ in the present model. The cancellation
is less significant if $\Gamma$ is larger in comparison with
$\nu_r$.

We turn now to numerical calculations of the conversion rates in $^{13}$CH$_3$F. 
Contribution from $m'-n'$ pair is difficult to calculate using 
Eq.~(\ref{g'}) directly because of some uncertainty in the parameters 
involved. Instead, one can use the experimental value,
$\gamma_{po}'=2.3\cdot10^{-3}$~s$^{-1}$/Torr
\cite{Nagels96PRL,Chap00AMR} and scale it linear in pressure. Such 
pressure dependence for $\gamma'_{po}$ is valid if 
$\Gamma\ll\omega'$ (see Eq.~(\ref{g'})) which is fulfilled for the pressures 
$P<10$~Torr.

Calculation of the rates $\gamma_{po}$, $\gamma_n$, and $\gamma_c$
needs the matrix elements $V_{mn}$. Mixing of the states $m$ and $n$ is 
performed by the intramolecular spin-spin interaction between the nuclei of 
$^{13}$CH$_3$F. Dependence of  $V_{mn}$ on nuclear spin variables is 
accounted easily by summation because other factors in Eqs.~(\ref{g})-(\ref{gc}) 
do not depend on nuclear spins. Then, the only remaining degeneracy is in 
$M$-quantum numbers. This quantity reads,
\begin{equation}
     \sum|V_{mn}|^2 = (2J_m+1)(2J_n+1)\left(\begin{array}{rcr}
         J_m  & 2 & J_n \\ 
        -K_m & q & K_n
       \end{array}\right)^2
       \left(\begin{array}{rcr}
         J_m  &    2 & J_n \\ 
          -M' & M'-M & M
       \end{array}\right)^2{\cal T}^2_{2,q}.
\label{vmn}
\end{equation}
Here ${\cal T}_{2,q}$ ($q=K_m-K_n=2$) is the magnitude of the spin-spin interaction
and summation is made in all nuclear spin projections.
The value of ${\cal T}_{2,2}$ calculated from the molecular structure
is equal to 69.2~kHz. This value is confirmed by the experiment
\cite{Cosleou00EPJD}. But in this paper we have to take somewhat smaller value, 
${\cal T}_{2,2}=64.1$~kHz, as it was obtained self-consistently for the three parameters,
${\cal T}_{2,2}$, $\Gamma$, and $\gamma'_{po}$ \cite{Chap00AMR}. 

Examples of the conversion rates are shown in Fig.~\ref{fig5}. The upper
panel gives the rate $\gamma$ for two pressures, 30~mTorr and 100~mTorr.
The radiation intensity in both cases is equal to 100~mW/cm$^2$.
One can see that the peak at $\Omega=-\omega_1$ is $\simeq$3 times larger
at 30~mTorr than at 100~mTorr. There is also peak at $\Omega=0$
having ``negative amplitude'' but it is too small to be visible in the
upper panel. The lower panel shows
the field dependent rates, $-(\gamma_n+\gamma_c)$. They are
taken with the same sign as they contribute to $\gamma$ in Eq.~(\ref{drp1}).  
One can see from this panel that the pressure dependences of the amplitudes
of these two peaks are opposite. This is the consequence of the crossing
of ortho and para states at $\Omega=-\omega_1$ and off-resonant nature of the peak at 
$\Omega=0$.

Broadening of the two peaks in the conversion rate are very different too.
The peak at $\Omega=0$ has the broadening as ordinary absorption line does.
At large saturation parameter, $\kappa$, its width is  $\sim2|G|$ and grows  
rather fast with intensity. The width of the peak at 
$\Omega=-\omega_1$ is given by $\simeq\Gamma(1+|G|^2/\omega^2)$. Consequently, the
power broadening of this peak is very small at our conditions. Fig.~\ref{fig5} 
illustrates the difference  in the peaks widths. 

\section{Enrichment}

Solution of the kinetic equation (\ref{drp1}) can be presented as,
\begin{equation}
     \rho_p(t) = \overline\rho_p + (\rho_p(0)-\overline\rho_p)\exp{(-\gamma t)};\ \ \ 
                 \overline\rho_p = N(\gamma'_{po}+\gamma_{po})/\gamma,
\label{rpt}
\end{equation}
where $\overline\rho_p$ and $\rho_p(0)$ are the steady-state and initial
(equilibrium) concentrations of para molecules, respectively. Enrichment of 
para molecules will be defined as,
\begin{equation}
     \beta(\Omega) = \frac{\overline\rho_p}{\rho_p(0)}-1.
\label{b}
\end{equation} 
Partition functions for ortho and para isomers of CH$_3$F are equal.
Consequently, $\gamma'_{po}=\gamma'_{op}$, $\gamma_{po}=\gamma_{op}$
and enrichment, $\beta(\Omega)$, can be expressed as,
\begin{equation}
     \beta(\Omega) = (\gamma_n+\gamma_c)/\gamma,
\label{bg}
\end{equation}
which will be used for the numerical calculations of $\beta(\Omega)$.
One can note from Eq.~(\ref{bg}) that despite the rates $\gamma_{po}$ and 
$\gamma_{op}$ have rather large field dependent parts (see Fig.~(\ref{fig5})), they alone 
would not produce an enrichment. It can be understood because these 
field dependent parts are due to the mixing of states shifted by radiation
but having equilibrium populations (see Eq.~(\ref{g})). Conversion through
the mixing of equilibrium populated states  
does not affect the ortho-to-para ratio. Examples of the enrichment $\beta(\Omega)$
are given in Fig.~\ref{fig6}. At the pressure 30~mTorr and radiation intensity
$S=100$~mW/cm$^2$, enrichment peak at $\Omega=-\omega_1$ is $\simeq$3\% and
it is $\simeq$4 times higher than the enrichment at $\Omega=0$. At
larger radiation intensity, amplitude of the  enrichment peak, 
$\beta(-\omega_1)$, saturates 
at $\simeq$5\%. The peak at $\Omega=0$ grows to $\simeq$1\%.

The simplified model that accounts only the radiation-induced level population
change predicts a factor two higher peak at $\Omega=0$ than the present
model. As was discussed above, in the present model one has 
partial cancellation of contributions
originated from the rates $\gamma_n$ and $\gamma_c$ which results in a
smaller enrichment. Another significant difference between these models is
that the simplified model does not predict an enrichment at  $\Omega=-\omega_1$.

Broadening of the two peaks in enrichment is very distinctive and is similar to the peaks
of conversion rate, although there is some difference. The enrichment peak
at $\Omega=-\omega_1$ is broader than the peak of $\gamma$ because
of large $\gamma_{po}$ and $\gamma_{op}$ at resonant frequency in the denominator of
Eq.~(\ref{bg}).

\section{Discussion and Conclusions}

We have developed a model of spin isomer coherent control that 
accounts for the molecular level degeneracy in magnetic quantum numbers. 
This degeneracy together with the
account of molecular center-of-mass motion allow now to apply the model 
to real molecules, having actual level structure and ortho-para mixing Hamiltonian.
The developed model was used to analyse the microwave induced enrichment of spin 
isomers in $^{13}$CH$_3$F. Spectrum of the conversion rate consists of two peaks 
with the peak at $\Omega=-\omega_1$ being two orders of magnitude higher than the
peak at $\Omega=0$.

The enrichment spectrum has also two peaks at the same frequencies. 
One can obtain 3\% enrichment using forbidden (for ordinary absorption) 
resonance at $\Omega=-\omega_1$, gas pressure 30~mTorr at room temperature 
and radiation intensity $S=100$~mW/cm$^2$. Amplitude of the peak at $\Omega=0$ 
is 4 times smaller. Cooling the gas to 200~K would increase enrichment
to $\simeq$5\% because of the increase of the level population difference.
At higher radiation intensity enrichment saturates at $\simeq$5\%
if the gas temperature $T=295$~K and at $\simeq$7\% if $T=200$~K.

For the practical implementation of the microwave enrichment, 
the use of the peak at $\Omega=-\omega_1$ has a few advantages. Enrichment here
is significantly larger than at $\Omega=0$. Moreover, spurious effects 
can decrease the enrichment at $\Omega=0$
even further. It can be due to the gas heating by radiation. Note, that
for the peak at $\Omega=-\omega_1$ absorption is negligible and there is
no gas heating. Another spurious affect may be due to the resonant
exchange of rotational quanta between ortho and para isomers. 
This effect prevents depopulation of one rotational state by radiation 
(state $n$ in our case) in comparison with the population of the state $m$ having 
nearly equal energy. Again, the peak at $\Omega=-\omega_1$ has the advantage
of very low absorption coefficient and thus low radiation induced population
change. Finally, the disadvantage of the peak at $\Omega=0$ is the spurious
absorption by the line $11,0\rightarrow12,0$ (Fig.~\ref{fig2}). Although at low
pressures this line is well separated from the line $11,1\rightarrow12,1$
(the gap is 10~MHz), power broadening partially overlap these lines. On
other hand, the peak $\Omega=-\omega_1$ is situated at 
the blue side of the line $11,0\rightarrow12,0$ being 120~MHz away from the
nearest absorption line (Fig.~\ref{fig2}).
 
Enrichment obtained by microwave excitation of $^{13}$CH$_3$F
at reasonable experimental conditions is not large, $\simeq$3\%.
On the other hand, there are some applications where such enrichment could be 
significant, e.g., in spin isomer enhanced NMR technique 
\cite{Bowers86PRL,Natterer97PNMRS}. Note that for the standard 200 MHz NMR
the difference in Boltzmann factors between Zeeman states,
which determines the amplitude of NMR signal, is only $3\cdot10^{-5}$.
If even a fraction of the 3\% isomer enrichment would be transported to the 
Zeeman level populations it would enhance the NMR signal significantly.

Apart from any possible applications of microwave enrichment, 
which is by far too early to discuss now,
observation of the microwave enrichment would have
prove-of-principle importance for the coherent
control of spin isomers. It is interesting to note also that the isomer enrichment at
$\Omega=-\omega_1$ would demonstrate an example of enhanced access to weak processes
in molecules through the isomer enrichment \cite{Chap01JPB}. 
Suppose, one would like to measure the absorption
at $\Omega=-\omega_1$ directly. At the conditions considered in the paper,
$\chi(-\omega_1)\simeq6\cdot10^{-6}$~cm$^{-1}$ which is rather
difficult to measure. This value has to be compared with the
3\% enrichment in the coherent control which should not be difficult to measure.

\section*{Acknowledgments}
The authors are indebted to K.A.Nasyrov for the useful discussions of light-molecule
interaction theory. This work was supported in part by the  
Russian Foundation for Basic Research (RFBR), grant No.~01-03-32905

\newpage
Table 1. Positions of levels in $^{13}$CH$_3$F. Molecular parameters are
from Ref.~\cite{Papousek94JMS}.

\vspace{3cm}
\begin{tabular}{|c|c|c|c|c|}
\hline
 Notation$^*$ & $J,K$ & $I$ & E(cm$^{-1}$) & Frequency (MHz) \\
\hline
 $m'$         & 20,3  & 3/2 & 387.1        & 351.01$\pm0.16$  ($m'-n'$)\\
 $n'$         & 21,1  & 1/2 & 387.1        &  \\
\hline
 $q$          & 12,1  & 1/2 & 133.7        &  596294.285$\pm$0.013  ($q-n$)\\
 $n$          & 11,1  & 1/2 & 113.8        &  130.99$\pm$0.15   ($n-m$)\\
 $m$          &  9,3  & 3/2 & 113.8        &  596425.28$\pm$0.15 ($q-m$)\\
\hline
\end{tabular}

\begin{figure}[hb]
\includegraphics[width=12cm]{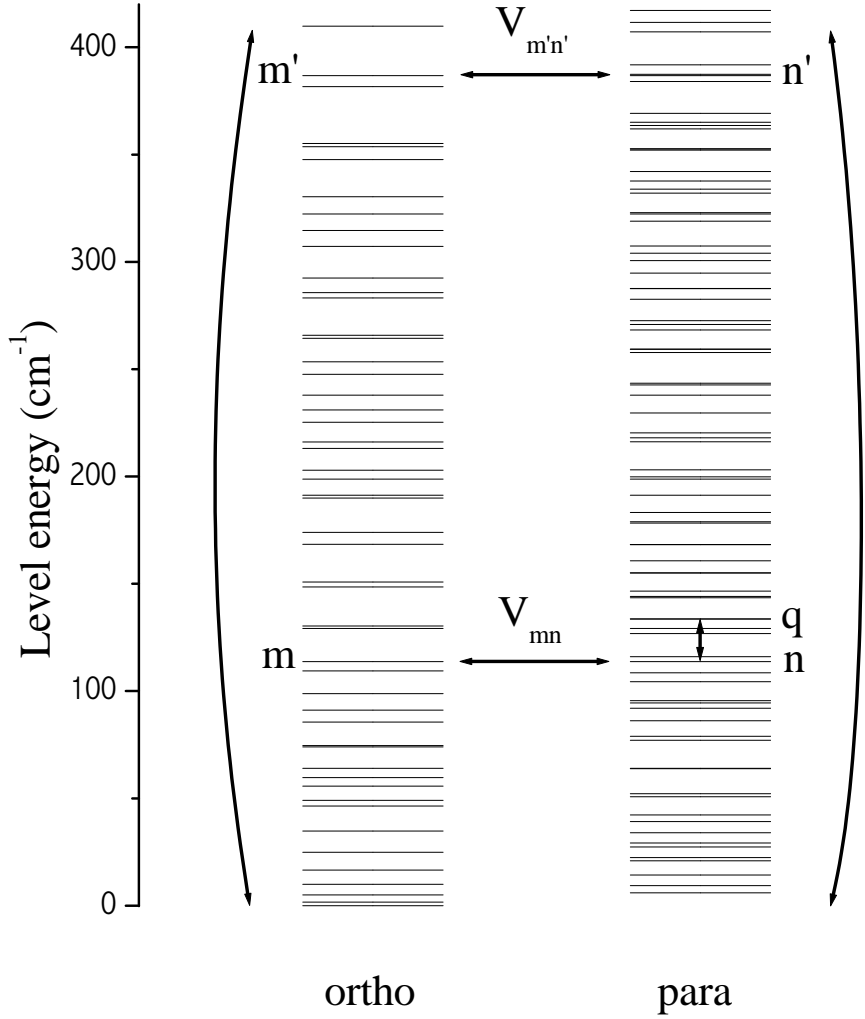}
\caption{Position of rotational levels of $^{13}$CH$_3$F.
Molecular parameters are from Ref.~\cite{Papousek94JMS}. 
Two pairs of states important for the spin conversion
in this molecule are indicated. Small vertical line in the para 
subspace indicates microwave
excitation of the transition $n\rightarrow q$. Two bent vertical 
lines indicate rotational relaxation. Parameters of the important
states are summarized in Table~1.}
\label{fig1}
\end{figure}

\begin{figure}[hb]
\includegraphics[width=16cm]{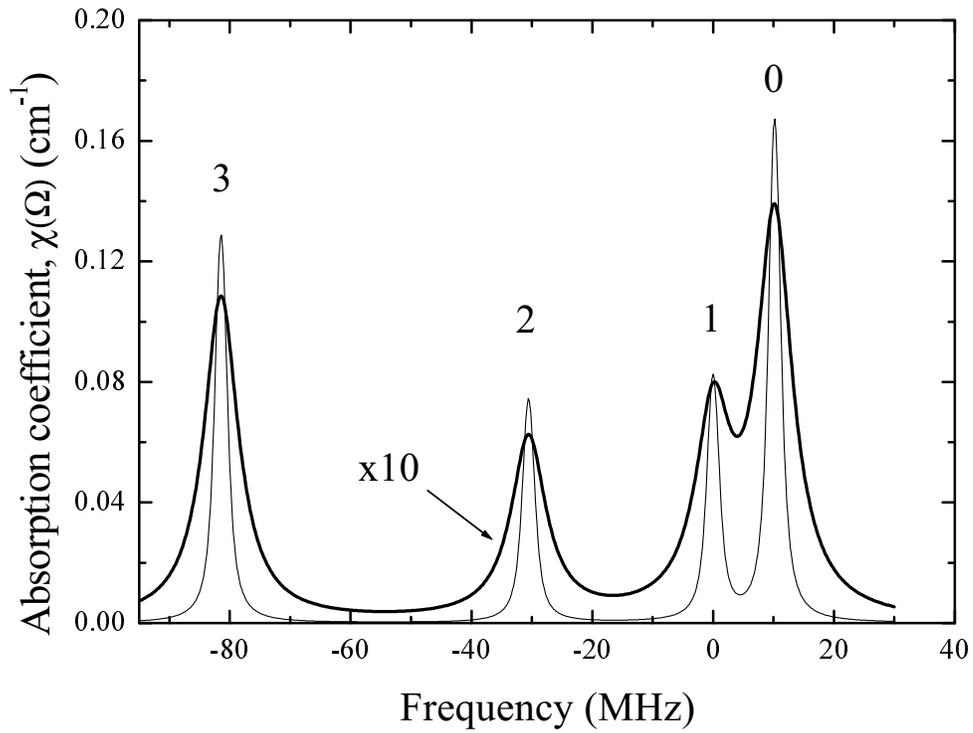}
\vspace{2cm}
\caption{Absorption spectrum near the line $11,1\rightarrow12,1$. 
Numbers in the graph indicate the $K$-values. The gas pressure is 
30~mTorr. Low intensity absorption is
shown by thin line, the case of $S=100$~mW/cm$^2$ is shown by thick line.}
\label{fig2}
\end{figure}

\begin{figure}[hb]
\includegraphics[width=18cm]{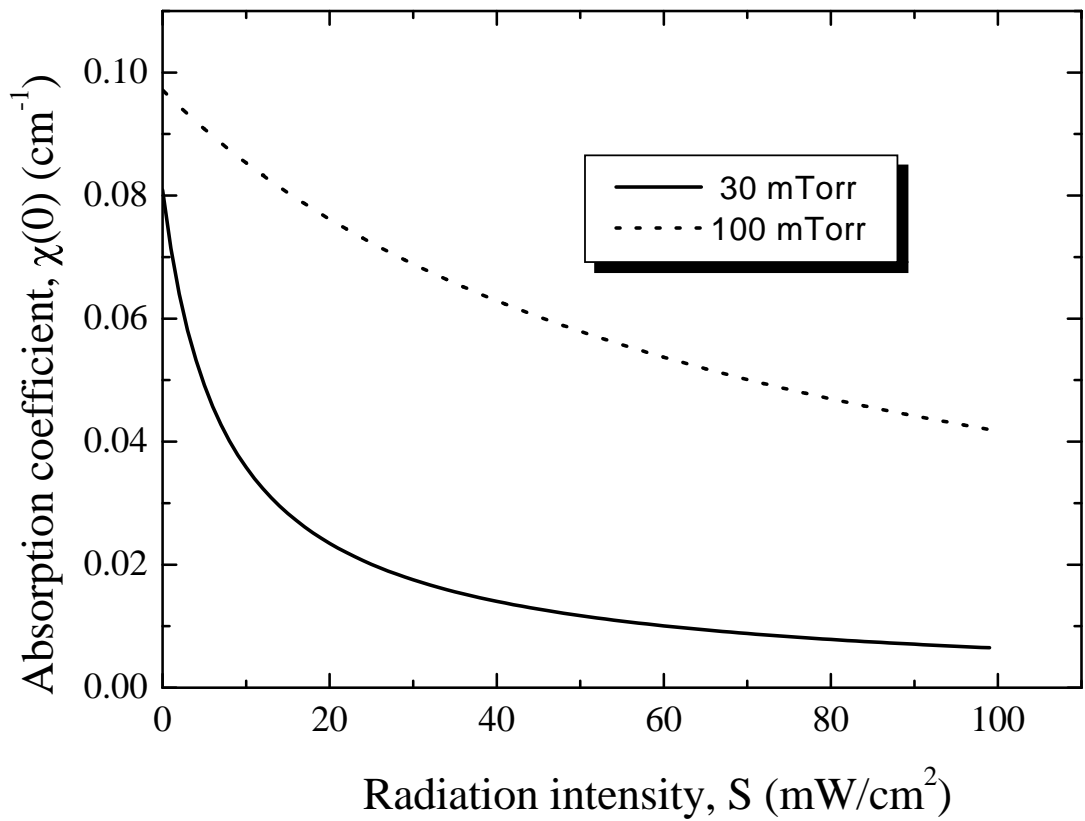}
\vspace{2cm}
\caption{Saturation of the absorption coefficient in the line center of
transition $11,1\rightarrow12,1$. Gas temperature is $T=295$~K.}
\label{fig3}
\end{figure}

\begin{figure}[hb]
\includegraphics[width=18cm]{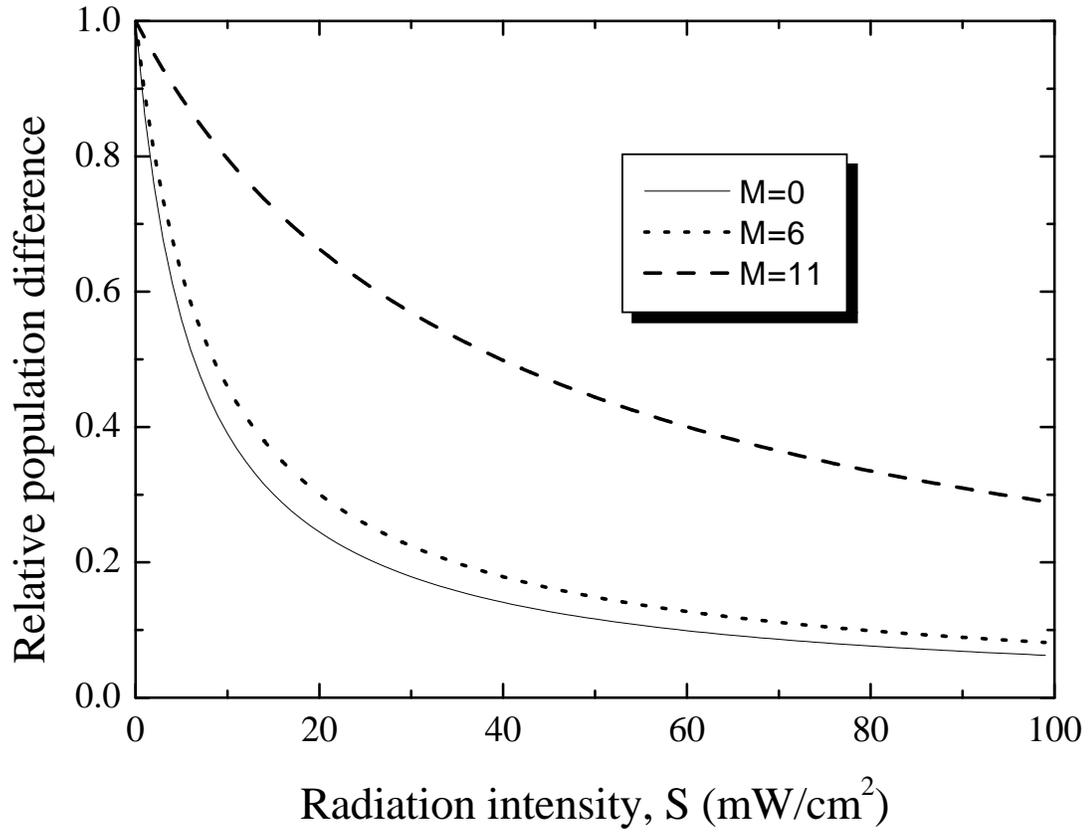}
\vspace{2cm}
\caption{Saturation of the level population difference
normalized to the field free population difference. Transition 
$11,1\rightarrow12,1$. Gas pressure, $P=30$~mTorr; gas temperature,
$T=295$~K.}
\label{fig4}
\end{figure}

\begin{figure}[hb]
\includegraphics[width=14cm]{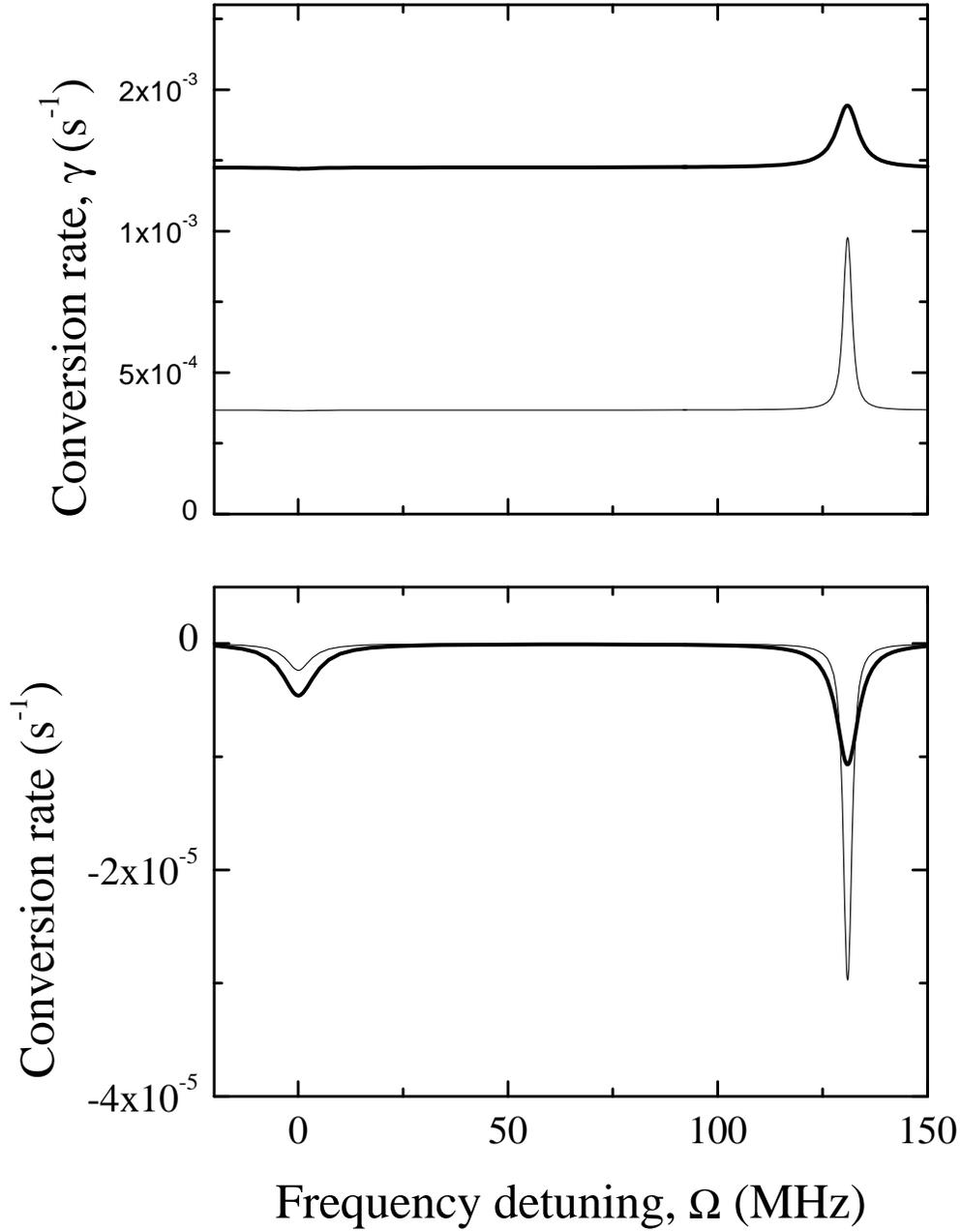}
\caption{Conversion rates. The upper panel shows the total conversion rate,
$\gamma$, for the pressures $P$=30~mTorr (thin line) and $P$=100~mTorr (thick line).
The lower panel shows the field dependent contribution, $-(\gamma_n+\gamma_c)$,
at the pressures $P$=30~mTorr (thin line) and $P$=100~mTorr (thick line). 
The radiation intensity is $S=100$~mW/cm$^2$}
\label{fig5}
\end{figure}

\begin{figure}[hb]
\includegraphics[width=18cm]{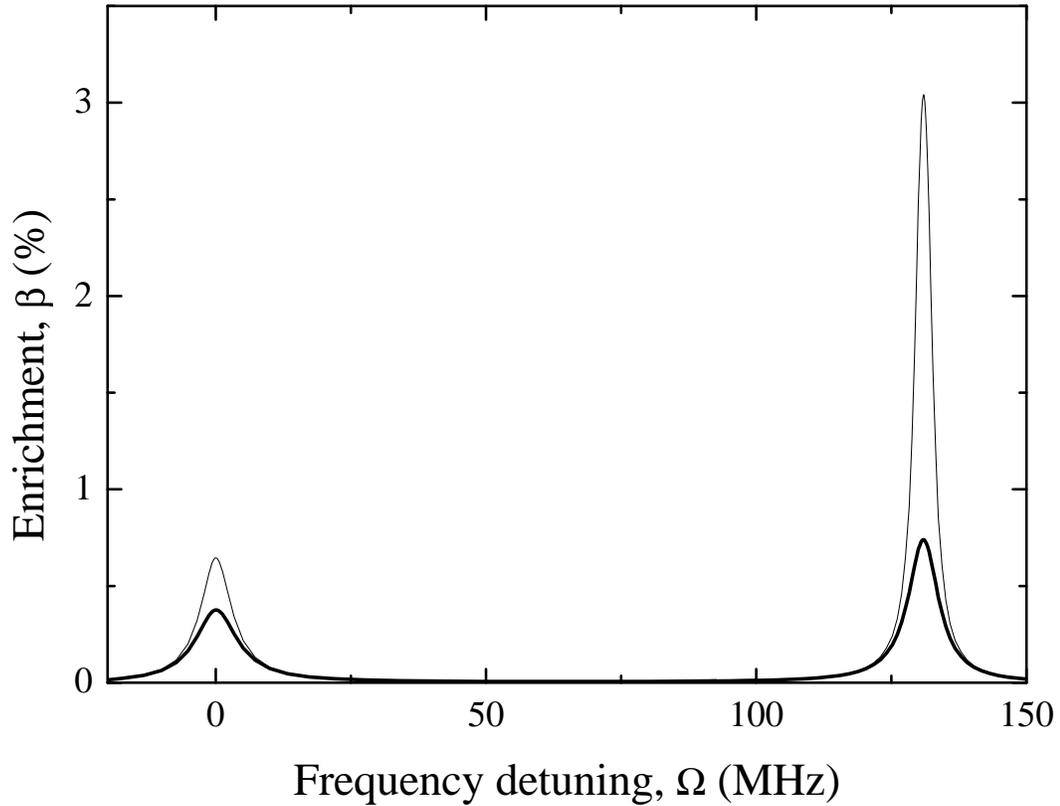}
\vspace{2cm}
\caption{Enrichment of para molecules, $\beta$, as a function of 
radiation frequency detuning, $\Omega$. Gas pressure, $P=30$~mTorr (thin line);
$P=100$~mTorr (thick line). In both cases the radiation intensity
$S=100$~mW/cm$^2$ and gas temperature $T=295$~K.}
\label{fig6}
\end{figure}

\vspace{1cm}
* Notation in Fig.~\ref{fig1}.

\newpage

\newpage

\newpage

\newpage

\newpage

\newpage

\end{document}